\documentclass[10pt, conference, compsocconf]{IEEEtran}

\usepackage[pdftex]{graphicx}
\usepackage{hyperref,pdfcomment}
\usepackage{url}
\usepackage{multirow}
\usepackage[table]{xcolor}

\let\labelindent\relax

\usepackage{enumitem}
\usepackage{subfigure}

\begin{document}

\title{Cloud Benchmarking for Performance}

\author{\IEEEauthorblockN{Blesson Varghese, Ozgur Akgun, Ian Miguel, Long Thai and Adam Barker}
\IEEEauthorblockA{School of Computer Science, University of St Andrews, UK\\
Email: \{varghese, ozgur.akgun, ijm, ltt2, adam.barker\}@st-andrews.ac.uk}
}

\maketitle

\begin{abstract}
\textit{How can applications be deployed on the cloud to achieve maximum performance?} This question has become significant and challenging with the availability of a wide variety of Virtual Machines (VMs) with different performance capabilities in the cloud. The above question is addressed by proposing a six step benchmarking methodology in which a user provides a set of four weights that indicate how important each of the following groups: memory, processor, computation and storage are to the application that needs to be executed on the cloud. The weights along with cloud benchmarking data are used to generate a ranking of VMs that can maximise performance of the application. The rankings are validated through an empirical analysis using two case study applications; the first is a financial risk application and the second is a molecular dynamics simulation, which are both representative of workloads that can benefit from execution on the cloud. Both case studies validate the feasibility of the methodology and highlight that maximum performance can be achieved on the cloud by selecting the top ranked VMs produced by the methodology. 
\end{abstract}

\IEEEpeerreviewmaketitle

\section{Introduction}
\label{introduction}
The cloud computing marketplace offers an assortment of on-demand resources with a wide range of performance capabilities. This makes it challenging for a user to make an informed choice as to which Virtual Machine (VM) needs to be selected in order to deploy an application for maximum performance. Often it is the case that users deploy their applications on an ad hoc basis, without understanding which VMs can provide maximum performance. This can result in the application under performing on the cloud, and consequently increasing running costs. The research presented in this paper aims to address this problem. 

One way to address the above problem is by benchmarking \cite{benchmark-1}. Benchmarks are often used to measure performance of computing resources and have previously been applied to cloud resources \cite{cloudbenchmark-1}. Benchmarking is usually performed independently of an application and does not take into account any bespoke requirements an application might have. 

We hypothesize that by taking into account the requirements of an application, along with cloud benchmarking data, VMs can be ranked in order of performance so that a user can deploy an application on a cloud VM, which will maximise performance. In this paper, maximum performance is defined as minimum execution time of an application.

In order to determine the VM that can maximise application performance on the cloud, we present a six step benchmarking methodology. All that the user provides as input is a set of four weights (from 0 to 5), which indicate how important each of the following groups: memory and process, local communication, computation and storage are to the underlying application. These weights are mapped onto each of the four groups, which are evaluated against all potential VMs that can host the application. The VMs are then ranked according to their performance. 

For the purposes of verifying our hypothesis, the methodology is validated by empirical analysis using two case studies; the first is used in financial risk and the second employed in molecular dynamics. The contributions of this paper are the development of a benchmarking methodology for selecting VMs that can maximise the performance of an application on the cloud, and the validation of the methodology against real world applications. 

The remainder of this paper is organised as follows. Section \ref{methodology} presents the cloud benchmarking methodology. Section \ref{cloudbenchmarks} considers the set up for gathering the benchmarks and presents the  benchmarks used in the methodology. Section \ref{validationstudy} considers two case study applications for validate the benchmarking methodology. Section \ref{conclusions} presents a discussion on related work and concludes this paper. 

\section{Cloud Benchmarking methodology}
\label{methodology}
\begin{figure*} [!ht]
	\centering
	\includegraphics[width = 0.87\textwidth]{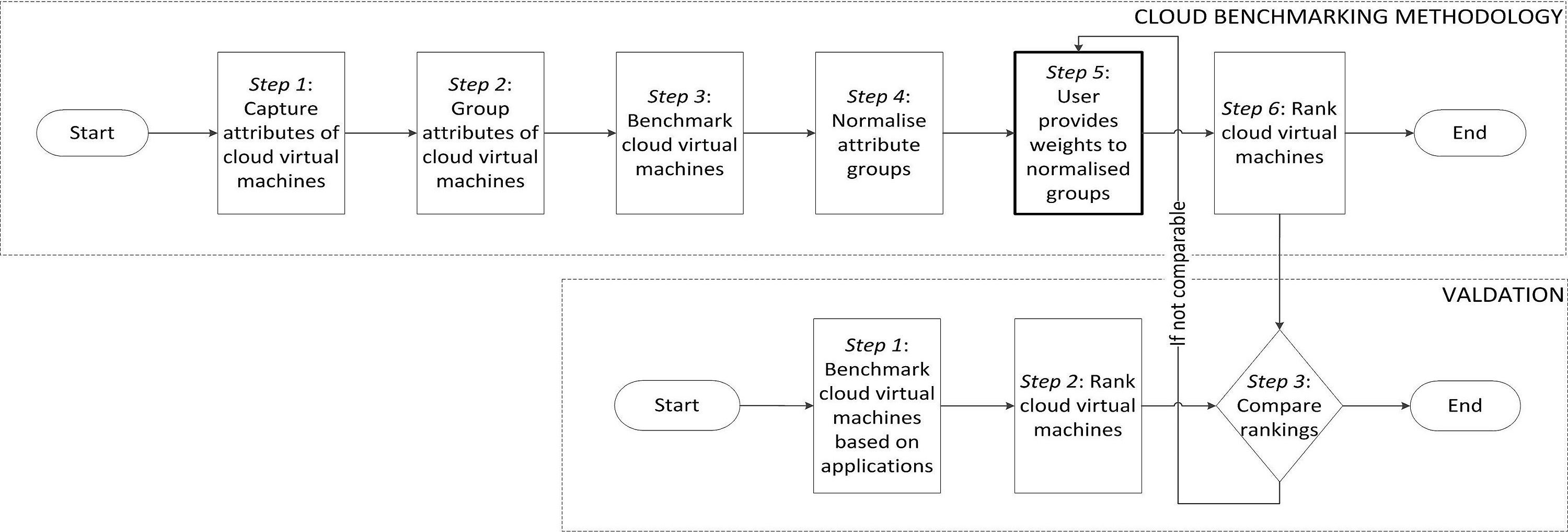}
	\caption{Cloud benchmarking methodology and validation technique}
	\label{figure0}
	\vspace{-0.5\baselineskip}
\end{figure*} 

The six step cloud benchmarking methodology is shown in Figure \ref{figure0}. 
Individual attributes of VMs are firstly evaluated and grouped together (Steps 1 to 4). The user provides a set of weights (or the order of importance) of the groups based on the requirements of the application (Step 5). The weights along with the grouped attributes for each VM are used to generate a score, which results in a performance-based VM ranking (Step 6). The VMs with highest ranks can maximise application performance. Consider there are $i = 1, 2, \cdots , m$ different VMs. The methodology is as follows:

\subsubsection*{Step 1: Capture Attributes}
Firstly, the attributes of a VM, which best describes it are selected. For example, attributes such as the number of integer, float and double addition, multiplication and division operations that can be performed in one second on a VM can describe its computational capacity. Consider there are $j = 1, 2, \cdots , n$ attributes of a VM. Then,  $r_{i,j}$ is the value associated with each $j^{th}$ attribute on the $i^{th}$ VM. 

\subsubsection*{Step 2: Group Attributes}
The attributes of the VM are grouped into categories based on whether they are related to memory and process, local communication or storage. For example, a group of attributes, such as the latencies of the main and random access memory and the L1 and L2 cache can be grouped as the memory group. Each attribute group is denoted as 
$G_{i, k} = \{r_{i, 1}, r_{i, 2}, \cdots \}$, where $i = 1, 2, \cdots m$, $k = 1, 2, \cdots , p$, and $p$ is the number of attribute groups.

\subsubsection*{Step 3: Benchmark Virtual Machines}
Based on the attribute groups a set of benchmarks are evaluated on all potential VMs. The benchmarks evaluate the attributes of the VM as closely as possible to the underlying hardware employed. Standard benchmarking tools are run on the VM or on an observer system to obtain the value of each attribute, $r_{i, j}$.

\subsubsection*{Step 4: Normalise Groups}
The attributes are normalised to rank the performance of VMs for an attribute group. 
The normalised value $\bar{r}_{i, j} = \frac{r_{i, j} - \mu_{j}}{\sigma_{j}}$, where $\mu_j$ is the mean value of attribute $r_{i, j}$ over $m$ VMs and $\sigma_j$ is the standard deviation of the attribute $r_{i, j}$ over $m$ VMs. The normalised attribute group, is denoted as $\bar{G}_{i, k} = \{\bar{r}_{i, 1}, \bar{r}_{i, 2}, \cdots \}$, where $i = 1, 2, \cdots m$, $k = 1, 2, \cdots , p$, and $p$ is the number of attribute groups.

\subsubsection*{Step 5: Weight Groups}
For a given application, some attribute groups may be more important than the others. For example, the file group is relevant for a simulation that has a large number of file read and write operations. This is known to domain experts who can provide a weight for each attribute group, which is defined as $W_{k}$. Each weight can take values from 0 to 5, where 0 indicates that the group is not relevant to the application and 5 indicates the importance of the group for the application.  

\subsubsection*{Step 6: Rank Virtual Machines}
The score of each VM is calculated as $S_{i} = \bar{G}_{i, k}.W_{k}$. The scores are ordered in a descending order for generating $Rp_{i}$ which is the ranking of the VMs based solely on performance. 

\subsection{Validating Cloud Benchmarking}
\label{validating}
An additional three steps are used for validating the cloud ranks generated by the methodology. They are as follows: 
\subsubsection*{Step 1: Application Benchmarking}
An application can be used for the empirical analysis by executing it on all VMs. The performance of the VMs is evaluated against a set of criteria (for example, the time taken by the application to complete execution). 

\subsubsection*{Step 2: Application Based Cloud Ranking}
The VMs are then ranked according to their empirical performance (in this paper performance ranking is with respect to the time taken for completing execution). 
The values of each criteria for evaluating performance are normalised using $\bar{v}_{i, j} = \frac{v_{i, j} - \mu_{j}}{\sigma_{j}}$, where $\mu_j$ is the mean value of $v_{i, j}$ over $m$ VMs and $\sigma_j$ is the standard deviation $v_{i, j}$ over $m$ VMs. The normalised values are used to rank the VMs $Mp_{i}$.

\subsubsection*{Step 3: Cloud Ranks Comparison}
$Rp_{i}$ from the methodology is compared against $Mp_{i}$. The rankings are comparable if the weights were properly chosen by the user. 
If there are significant differences between the rankings then the application requirements need to be re-evaluated and different weights assigned to the attribute groups.

\section{Cloud Benchmarks}
\label{cloudbenchmarks}
The experimental setup for obtaining the attributes of VMs by benchmarking and then grouping the attributes are presented in this section. The Amazon Elastic Compute Cloud (EC2)\footnote{\url{http://aws.amazon.com/ec2/}} is a rapidly growing public cloud infrastructure offering a variety of VMs with different performance capabilities. Hence, EC2 is the platform chosen for this research. Table \ref{table1} shows the VMs considered in this paper. 

\begin{table}[ht]
	\caption{Amazon EC2 VMs employed for benchmarking. \vspace{-2\baselineskip}
	}
	\label{table1}
\begin{center}
	\begin{tabular}{c c p{0.7cm} p{2.3cm} p{0.9cm}}
		\hline	
		\textbf{VM Type}	&	\textbf{vCPUs}	&	\textbf{Memory (GiB)}	&	\textbf{Processor}	& \textbf{Clock (GHz)}	\\
		\hline	
		\hline	
		\texttt{m1.xlarge}	&	4	&	15.0	&	Intel Xeon E5-2650	&	2.00\\
		\hline
		\texttt{m2.xlarge}	&	2	&	17.1	&	Intel Xeon E5-2665	&	2.40\\
		\texttt{m2.2xlarge}	&	4	&	34.2	&	Intel Xeon E5-2665	&	2.40\\
		\texttt{m2.4xlarge}	&	8	&	68.4	&	Intel Xeon E5-2665	&	2.40\\
		\hline
		\texttt{m3.xlarge}	&	4	&	15.0	&	Intel Xeon E5-2670	&	2.60\\
		\texttt{m3.2xlarge}	&	8	&	30.0	&	Intel Xeon E5-2670	&	2.60\\
		\hline
		\texttt{cr1.8xlarge}	&	32	&	244.0	&	Intel Xeon E5-2670	&	2.60\\		
		\hline
		\texttt{cc1.4xlarge}	&	16	&	23.0	&	Intel Xeon X5570	&	2.93\\
		\texttt{cc2.8xlarge}	&	32	&	60.5	&	Intel Xeon X5570	&	2.93\\
		\hline		
		\texttt{hi1.4xlarge}	&	16	&	60.5	&	Intel Xeon E5620	&	2.40\\
		\texttt{hs1.8xlarge}&	16	&	117.0	&	Intel Xeon E5-2650	&	2.00\\
		\hline
		\texttt{cg1.4xlarge}&	16	&	22.5	&	Intel Xeon X5570	&	2.93\\
		\hline		
	\end{tabular}
	\end{center}
	\vspace{-1\baselineskip}
\end{table}

\begin{figure*}[htp]
\begin{center}
	\subfigure[Memory Latencies: L1 and L2 cache] {\includegraphics[width=0.485\textwidth]{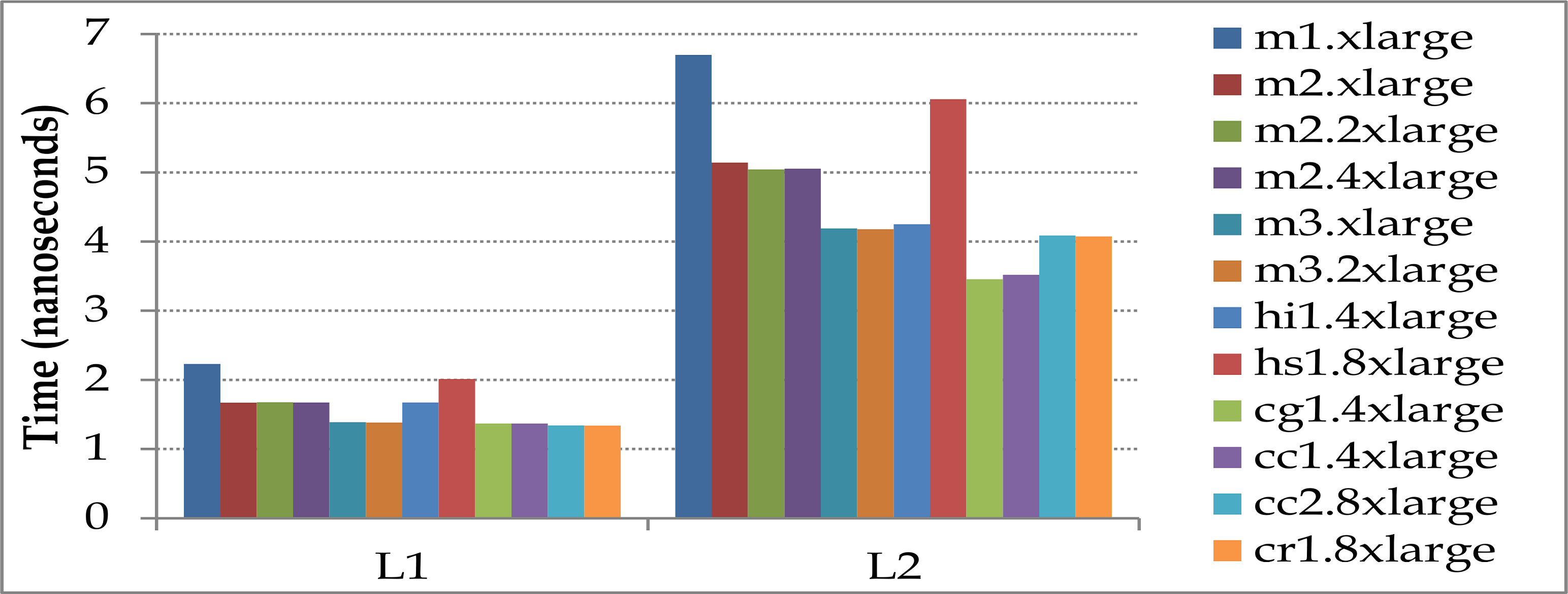} \label{figure3-1-1}}
	\subfigure[Memory Latencies: Main and Random Memory]{\includegraphics[width=0.485\textwidth]{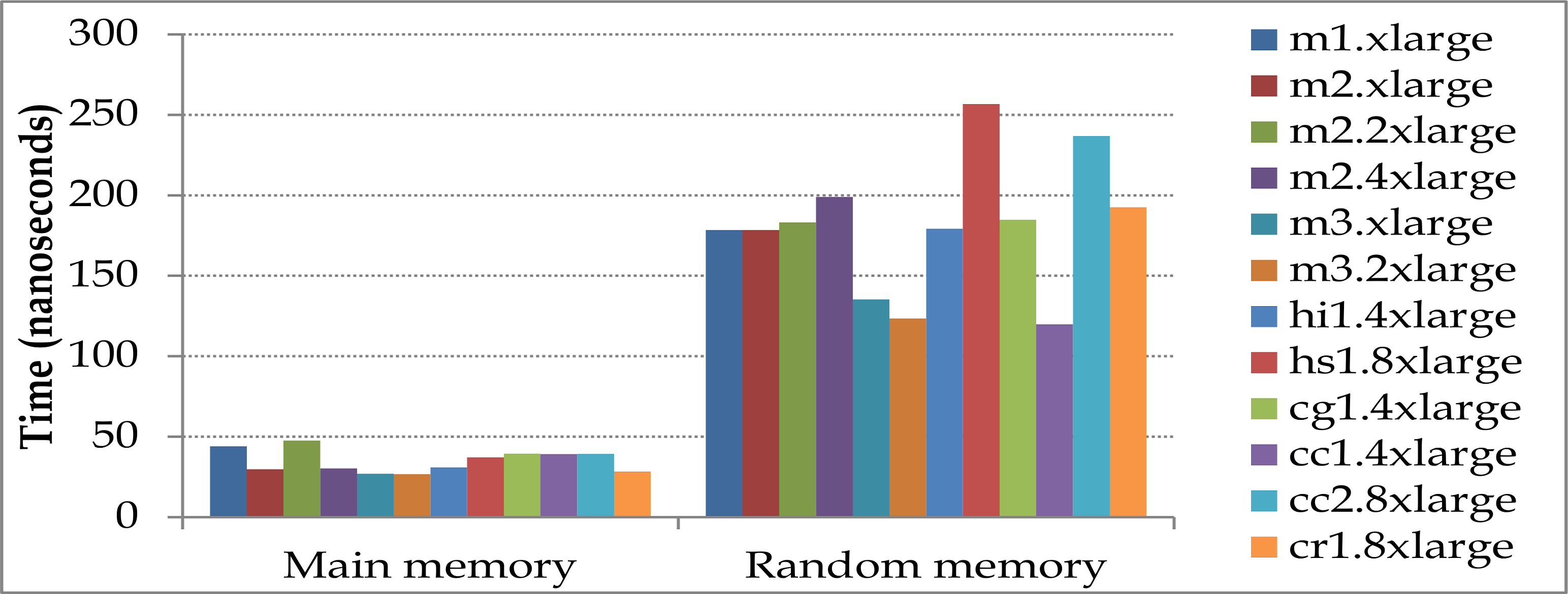} \label{figure3-1-2}}

	\subfigure[Local communication bandwidth]{\includegraphics[width=0.99\textwidth]{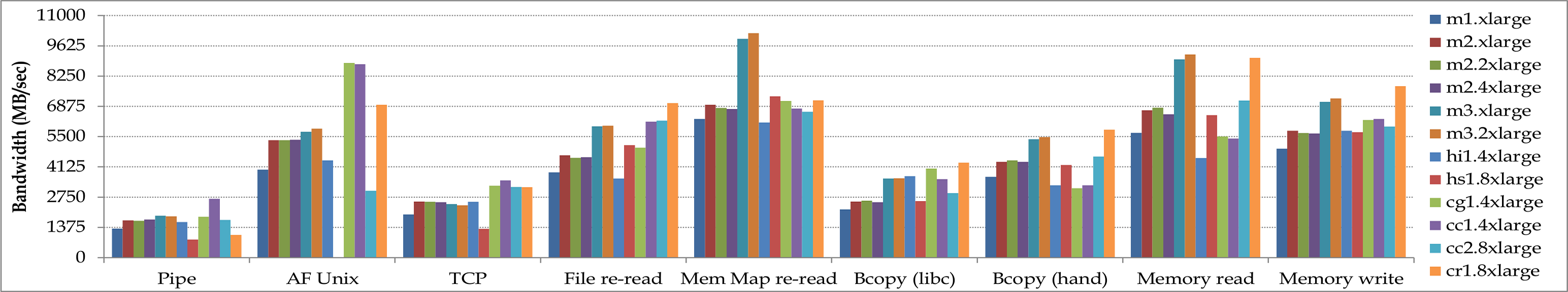} \label{figure3-2-1}}

	\subfigure[Arithmetic Operation Time: Addition and multiplication]{\includegraphics[width=0.535\textwidth]{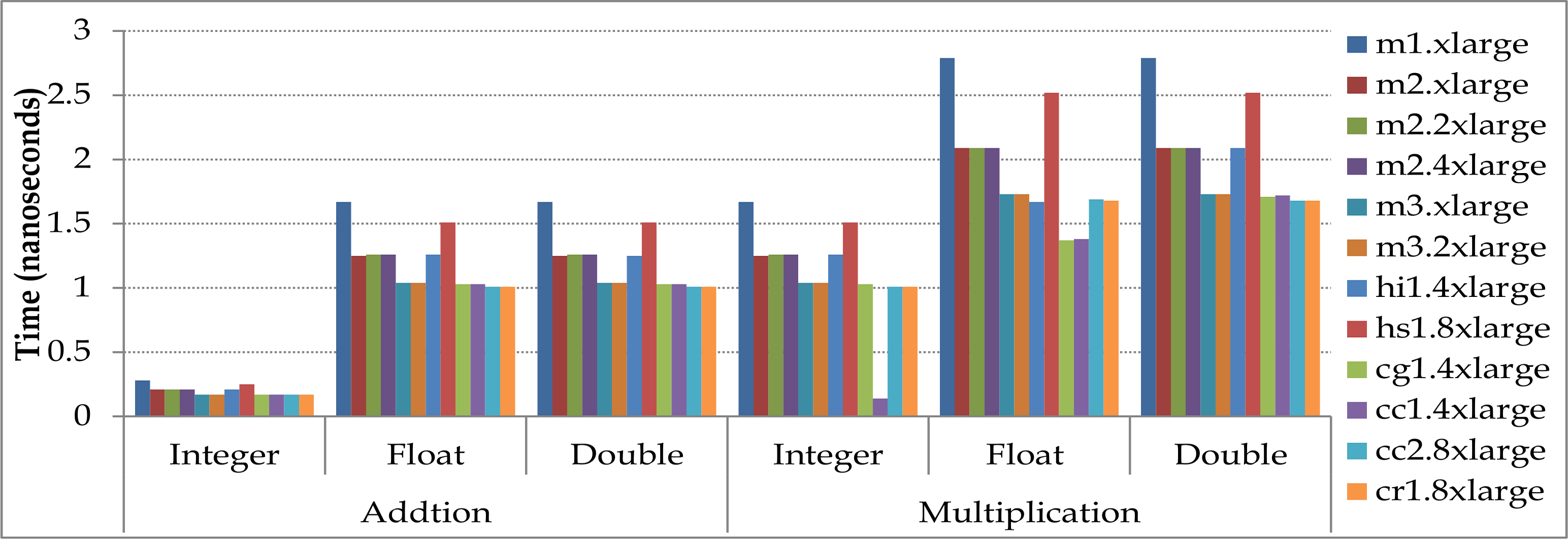} \label{figure3-3-1}}
	\subfigure[Arithmetic Operation Time: Division and Modulus]{\includegraphics[width=0.435\textwidth]{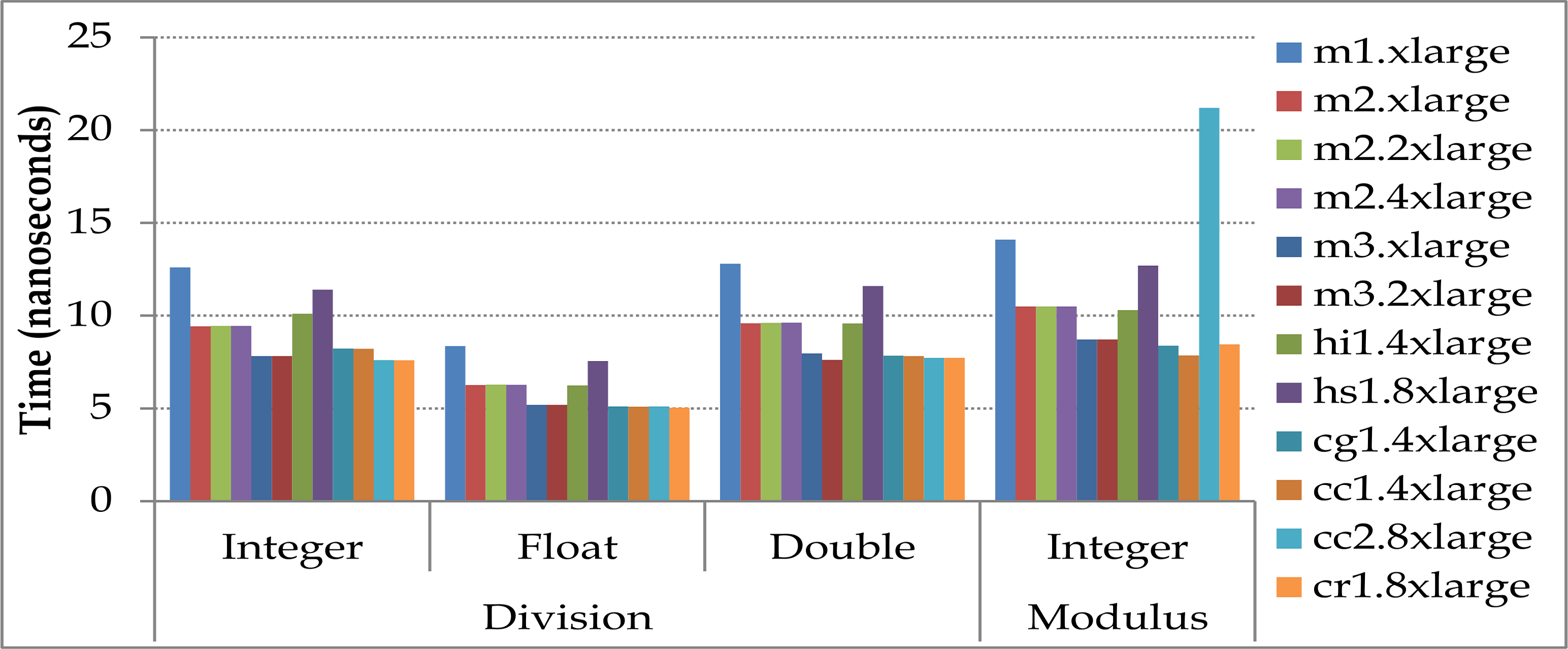} \label{figure3-3-2}}

	\subfigure[File I/O Operations: Sequential and random create and delete]{\includegraphics[width=0.61\textwidth]{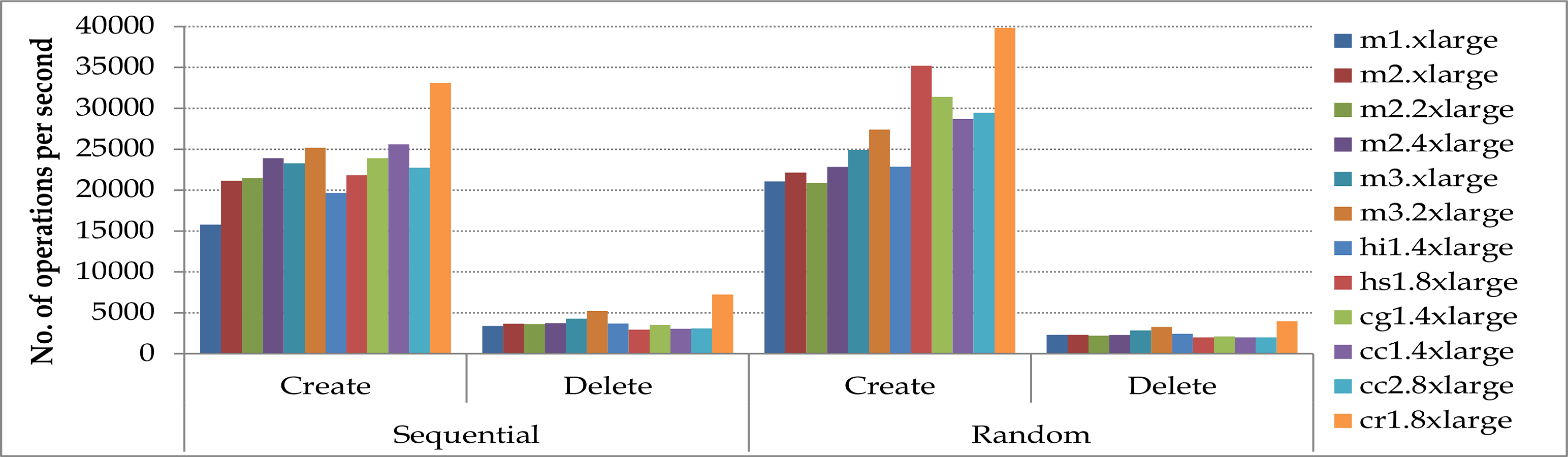} \label{figure3-4-1}}
	\subfigure[File I/O Operations: Sequential and random read]{\includegraphics[width=0.36\textwidth]{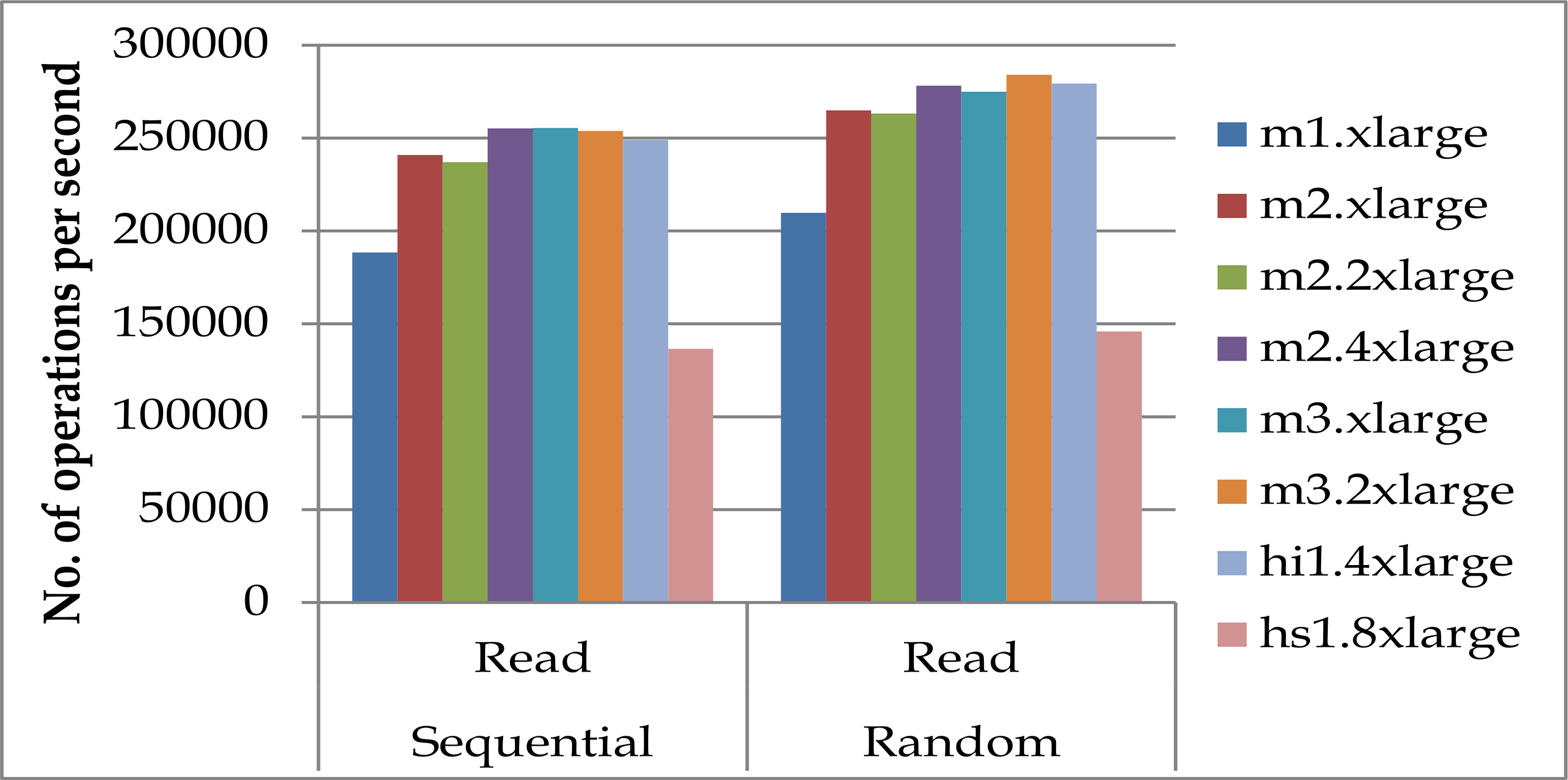} \label{figure3-4-2}}
\caption{Benchmarking on Cloud Virtual Machines}
\end{center}
\vspace{-1\baselineskip}
\end{figure*}

Three tools, namely (i) \texttt{bonnie++}\footnote{\url{http://sourceforge.net/projects/bonnie/}}
is used for file system benchmarks, (ii) \texttt{lmbench}\footnote{\url{http://lmbench.sourceforge.net/}} tool can provide latency and bandwidth information on top of a wide range of memory and process related information, and (iii) \texttt{sysbench}\footnote{\url{http://sysbench.sourceforge.net/}} tool, commonly referred to as the Multi-threaded System Evaluation Benchmark, is also used for obtaining benchmark metrics related to the CPU and the file I/O performance under data intensive loads. All experiments to gather the VM attributes were performed between six to eight times.

The attributes $r_{i, j}$ considered in Section \ref{methodology} are obtained using the above tools and then grouped to obtain $G_{i, k}$. The following four attribute groups are employed:

\subsubsection{Memory and Process Group}
This group, denoted as $G_{1}$ is used to benchmark the performance and latencies of the processor. Main memory and random memory latencies of the VMs are shown in Figure \ref{figure3-1-1} and the L1 and L2 cache latencies are shown in Figure \ref{figure3-1-2}.

\subsubsection{Local Communication Group}
The bandwidth of both memory communications and interprocess communications are captured under the local communication group, denoted as $G_{2}$. Figure \ref{figure3-2-1} shows memory communication metrics, namely the rate (MB/sec) at which data can be read from and written to memory, and interprocess communication metrics, namely the rate of data transfer between Unix pipes, AF\_Unix sockets and TCP.

\subsubsection{Computation Group}
The attributes captured in this group, denoted as $G_{3}$, are for benchmarking the performance of integer, float and double operations such as addition, multiplication and division and modulus. The time taken by the VMs for one addition and one multiplication operation performed using integer, float and double is shown in Figure \ref{figure3-3-1} and the time taken for one integer, float and double division operation and one integer modulus operation is shown in Figure \ref{figure3-3-2}. 

\subsubsection{Storage Group}
File I/O related attributes are grouped as the storage group, denoted as $G_{4}$, and considers sequential create, read and delete and random create, read and delete operations shown in Figure \ref{figure3-4-1} and Figure \ref{figure3-4-2}.

\section{Validation Study}
\label{validationstudy}
A user can provide a set of weights $W=\{W_{1}, W_{2}, W_{3}, W_{4}\}$ corresponding to each group and its importance for an application. Each weight takes values between 0 and 5, where 0 signifies that the group has no relevance to the application and 5 indicates the group is important for achieving good performance. All possible rankings for different weight combinations are generated. There are $6^4 - 1 = 1295$ (four groups and six values ranging from 0 to 5 for each weight, and discarding $W = \{0, 0, 0, 0\}$, which has no real significance) combinations of weights. The number of virtual CPUs is taken into account for parallel execution. 

\begin{figure}
\centering
	\subfigure[Sequential execution]{\label{figure50a}\includegraphics[width=0.44\textwidth]{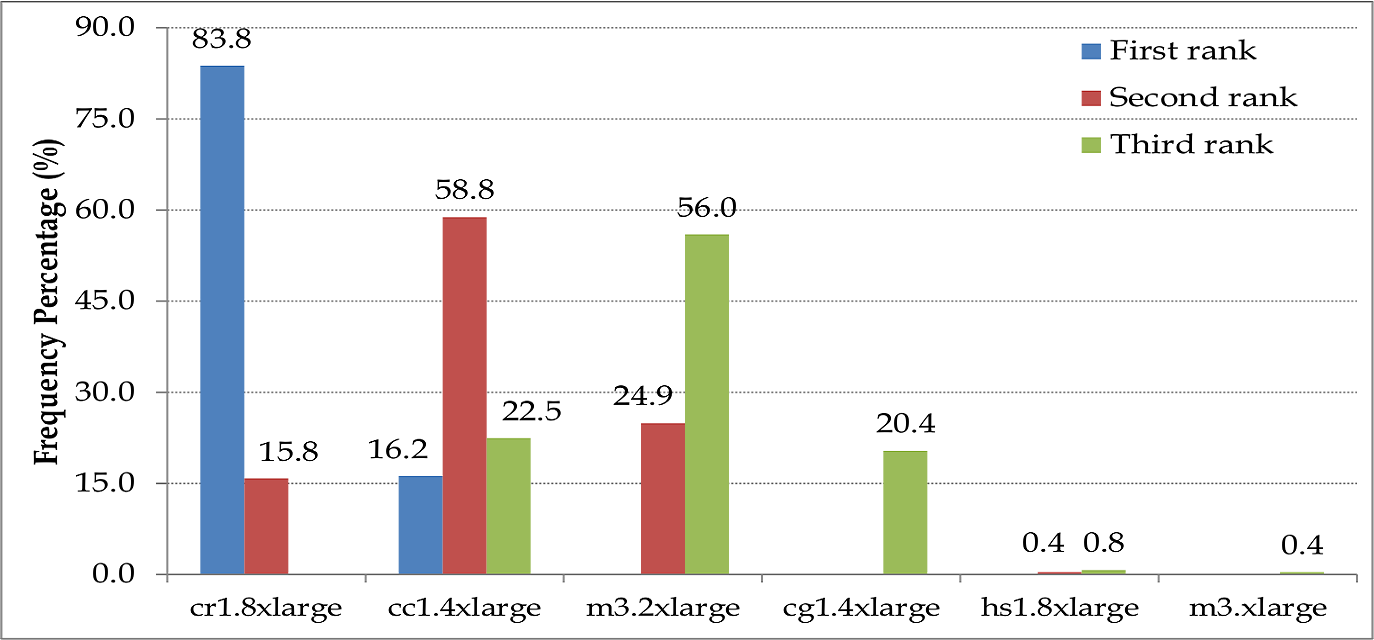}} \\
	\subfigure[Parallel execution]{\label{figure51a}\includegraphics[width=0.44\textwidth]{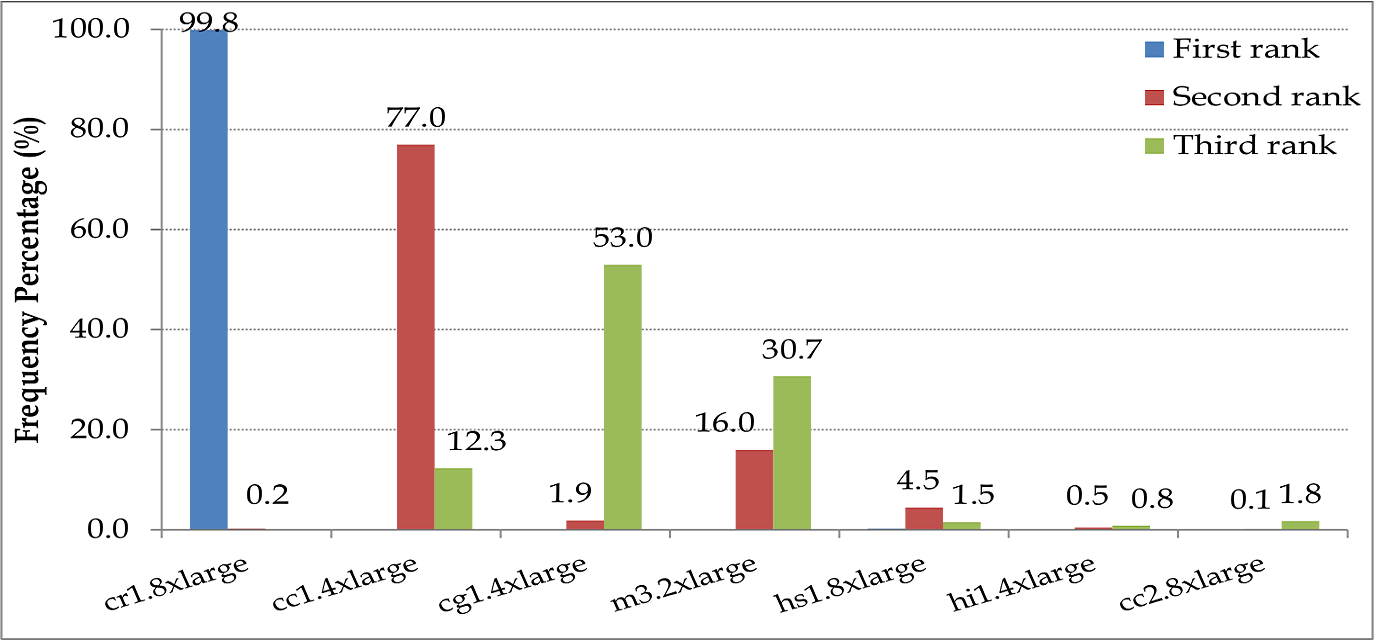}} 
\caption{Frequency of VMs that appear in top three ranks for all weights}
\label{figure50}
\vspace{-1\baselineskip}
\end{figure}

Figure \ref{figure50a} shows the six VMs that appear in the top three ranks for all possible weight combinations under sequential execution. The \texttt{cr1}, \texttt{cc1} and \texttt{m3.2xlarge} instances appear in the top three ranks with nearly 84\%, 59\% and 56\% frequency for all combination of weights respectively.

Seven VMs appear in the top three ranks for all possible weight combinations when parallel execution is taken into account (refer Figure \ref{figure51a}). Similar to sequential execution, the \texttt{cr1} instance dominates the first position followed by \texttt{cc1} which occupies the second rank in 77\% cases. The third rank is shared between \texttt{cg1} over 50\% and \texttt{m3.2xlarge} over 30\%. The key observation here is that there are clear winners who dominate the top three ranks when only performance is taken into account.

\subsection{Case Studies} 
The benchmarking methodology is validated on two case study applications. The applications are representative of different workloads that can benefit from using the cloud. The first application is used in the financial risk industry, referred to as Aggregate Risk Analysis \cite{agganalysis-1}. The application simulates a million trials of earthquake events and manages portfolios of insurance contracts in a year. The loss of each event is computed and used for real-time pricing.  

The second application is a molecular dynamics simulation \cite{md-1} of short range interactions used by theoretical physicists of a 10,000 particle system. The application computes the trajectory of the particles and the forces they exert on each other using a system of differential equations discretized into 200 time steps. For each time step the position and velocity of the particle are computed. 

\subsection{Benchmarking the Case Studies}
Based on the advice from industry experts and practitioners the weights for the first case study application is set as $W = \{5, 3, 5, 0 \}$. The application is memory and computationally intensive; local communication and file operations are less relevant. The top three VMs that achieve maximum performance are \texttt{cc1}, \texttt{cr1} and \texttt{cg1} shown in green in Table \ref{tablecasestudy1}.

\begin{table}
	\centering
	\caption{Ranking of VMs for sequential and parallel performance in the aggregate risk analysis case study ($W=\{5, 3, 5, 0\}$)}
	\begin{tabular}{| p{2cm} | p{1.1cm} | p{1cm} | p{1.1cm} | p{1cm} |}
		\hline
		\multirow{2}{*}{Amazon VM}	&	\multicolumn{2}{c |}{Sequential Ranking}	& \multicolumn{2}{c |}{Parallel Ranking}\\
		\cline{2-5}
		&	Benchmark &	Empirical Analysis	&	Benchmark 	&	Empirical Analysis\\
		\hline
		\hline
		\texttt{m1.xlarge}	&	12	&	12	&	12	&	10\\	
		\texttt{m2.xlarge}	&	8	&	8	&	10	&	12\\
		\texttt{m2.2xlarge}	&	10	&	8	&	9	&	9\\
		\texttt{m2.4xlarge} &	7	&	8	&	8	&	7\\
		\texttt{m3.xlarge}	&	5	&	5	&	6	&	11\\
		\texttt{m3.2xlarge}	&	4	& 	5	&	5	&	8\\
		\texttt{cr1.8xlarge}	&	\cellcolor{green!25}2	&	\cellcolor{blue!25}1	&	\cellcolor{green!25}1	&	\cellcolor{blue!25}2\\
		\texttt{cc1.4xlarge}	&	\cellcolor{green!25}1	&	\cellcolor{blue!25}2	&	\cellcolor{green!25}2	&	4\\
		\texttt{cc2.8xlarge}	&	6	&	\cellcolor{blue!25}2	&	4	&	\cellcolor{blue!25}1\\
		\texttt{hi1.4xlarge}	&	9	&	9	&	7	&	6\\
		\texttt{hs1.8xlarge}	&	11	&	11	&	11	&	\cellcolor{blue!25}3\\
		\texttt{cg1.4xlarge}	&	\cellcolor{green!25}3	&	\cellcolor{blue!25}2	&	\cellcolor{green!25}3	&	4\\
		\hline
	\end{tabular}
	\label{tablecasestudy1}
\end{table}

The second case study application is computationally intensive followed by the memory and processor requirements along with the need for local communication. There are no file intensive operations in this simulation. In consultation with domain scientists we set $W=\{4, 3, 5, 0\}$. The top three VMs that achieve maximum sequential and parallel performance are shown in Table \ref{tablecasestudy2} in green. Sequential performance can be maximised on \texttt{cc1.4xlarge}, \texttt{cr1.8xlarge} and \texttt{m3.2xlarge} VMs. Parallel performance is maximised on \texttt{cr1.8xlarge}, \texttt{cc2.8xlarge} and \texttt{cc1.4xlarge} VMs. 

\begin{table}
	\centering
	\caption{Ranking of VMs for sequential and parallel performance in the molecular dynamics case study ($W=\{4, 3, 5, 0\}$)}
	\begin{tabular}{| p{2cm} | p{1.1cm} | p{1cm} | p{1.1cm} | p{1cm} |}
		\hline
		\multirow{2}{*}{Amazon VM}	&	\multicolumn{2}{c |}{Sequential Ranking}	& \multicolumn{2}{c |}{Parallel Ranking}\\
		\cline{2-5}
		&	Benchmark &	Empirical Analysis	&	Benchmark &	Empirical Analysis\\
		\hline
		\hline
		\texttt{m1.xlarge}	&	11	&	10	&	11	&	10\\	
		\texttt{m2.xlarge}	&	9	&	8	&	9	&	11\\
		\texttt{m2.2xlarge}	&	7	&	7	&	8	&	8\\
		\texttt{m2.4xlarge} &	6	&	6	&	7	&	6\\
		\texttt{m3.xlarge}	&	4	&	5	&	5	&	9\\
		\texttt{m3.2xlarge}	&	\cellcolor{green!25}3	& 	\cellcolor{blue!25}3	&	4	&	7\\
		\texttt{hi1.4xlarge}&	8	&	9	&	6	&	4\\
		\texttt{hs1.8xlarge}&	10	&	11	&	10	&	5\\
		\texttt{cc1.4xlarge}&	\cellcolor{green!25}1	&	4	&	\cellcolor{green!25}2	&	\cellcolor{blue!25}3\\
		\texttt{cc2.8xlarge}&	5	&	\cellcolor{blue!25}2	&	\cellcolor{green!25}3	&	\cellcolor{blue!25}2\\
		\texttt{cr1.8xlarge}&	\cellcolor{green!25}2	&	\cellcolor{blue!25}1	&	\cellcolor{green!25}1	&	\cellcolor{blue!25}1\\
		\hline
	\end{tabular}
	\label{tablecasestudy2}
	\vspace{-0.5\baselineskip}
\end{table}

\subsection{Empirical Analysis of the Case Studies}
The applications were executed on the VMs both in sequence and in parallel to empirically verify the ranks. The top three ranks are shown in blue in Table \ref{tablecasestudy1} and Table \ref{tablecasestudy2} 

\begin{figure} 
\begin{center}
\label{figure4}
\subfigure[Sequential performance]{\includegraphics[width=0.44\textwidth]{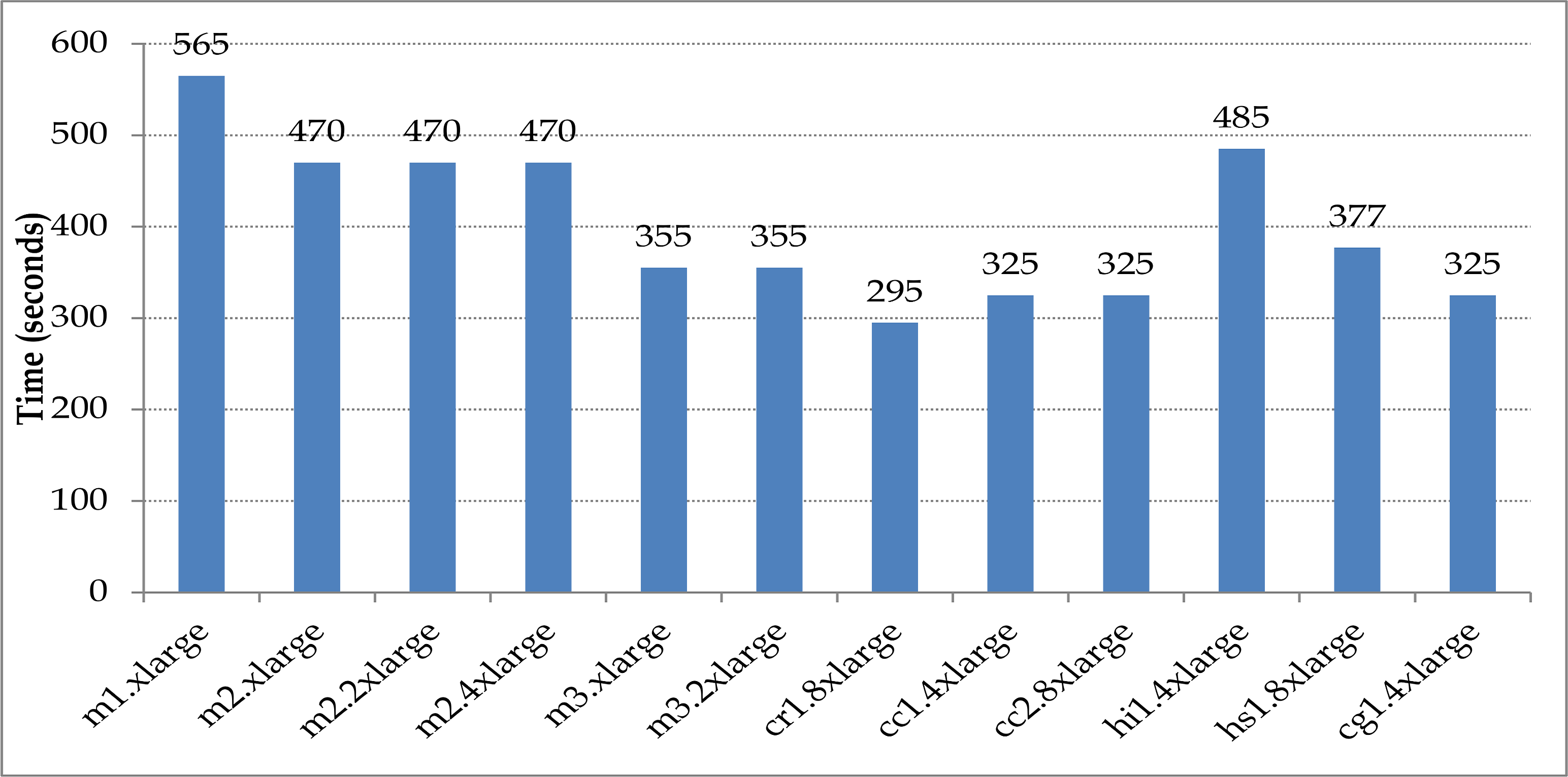} \label{figure4-1}}\hfill
\subfigure[Parallel performance on 16 threads]{\includegraphics[width=0.44\textwidth]{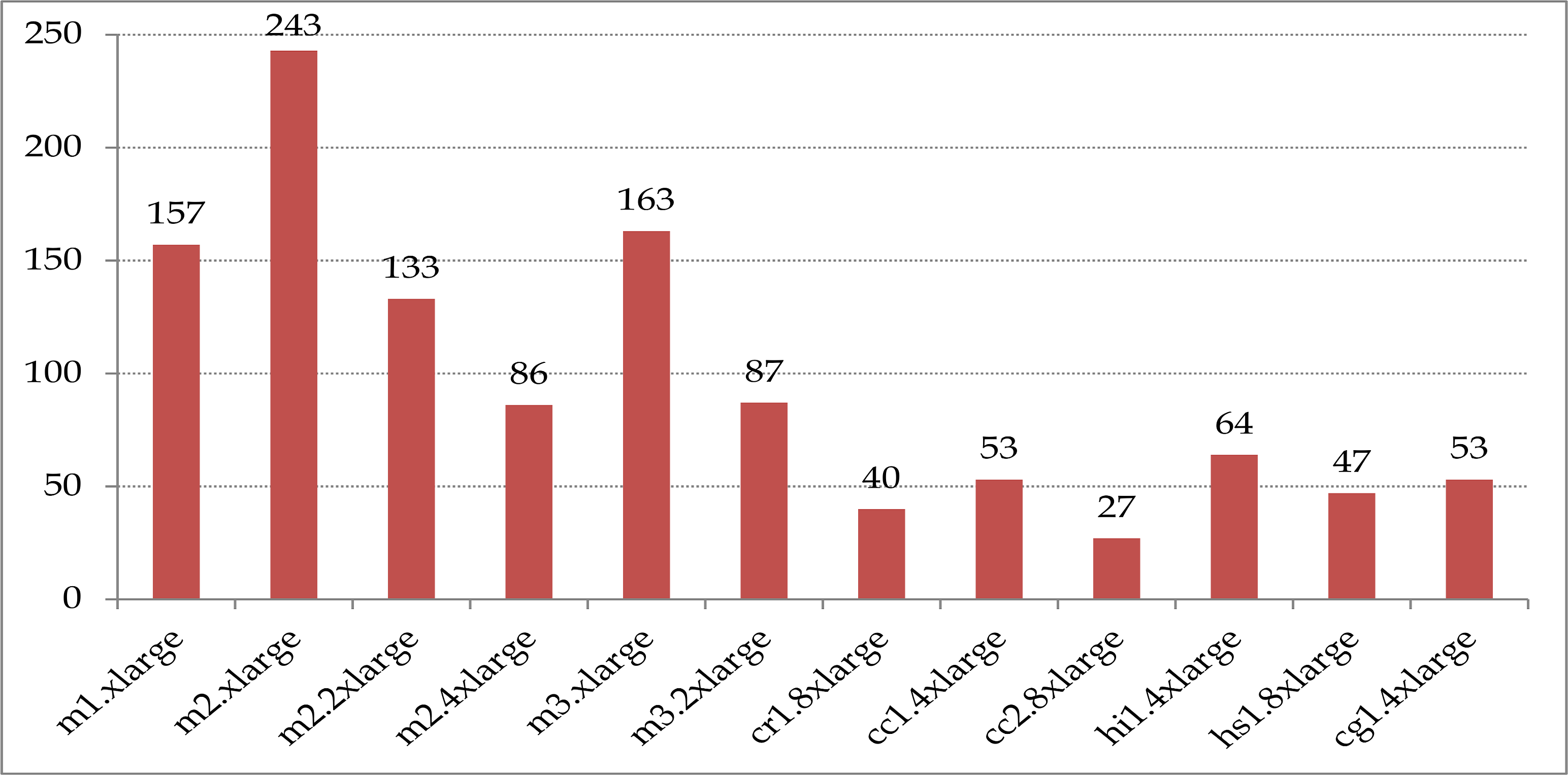} \label{figure4-2}}
\caption{Empirical study of the aggregate risk analysis case study}
\end{center}
\vspace{-1\baselineskip}
\end{figure}

Figure \ref{figure4-1} and Figure \ref{figure4-2} shows the time taken for performing Aggregate Risk Analysis on the Amazon instances sequentially (on one thread) and in parallel (on 16 threads) respectively. Sequential performance varies from a low of 565 seconds on \texttt{m1.xlarge} to a maximum of 295 seconds on the \texttt{cr1} instance; surprisingly, there is up to a 200\% difference in performance. As expected the parallel execution yields a speed up; a speed up of nearly 9x is obtained in the best parallel implementation on \texttt{cc2} over the baseline implementation on \texttt{cr1}. Although the \texttt{cr1} and \texttt{cc2} instances offer 32 virtual CPUs, the former achieves an acceleration of nearly 8.5x and the latter a speed up of 10x over baseline implementations. 

\begin{figure} 
\begin{center}
\label{figure6}
\subfigure[Sequential performance]{\includegraphics[width=0.44\textwidth]{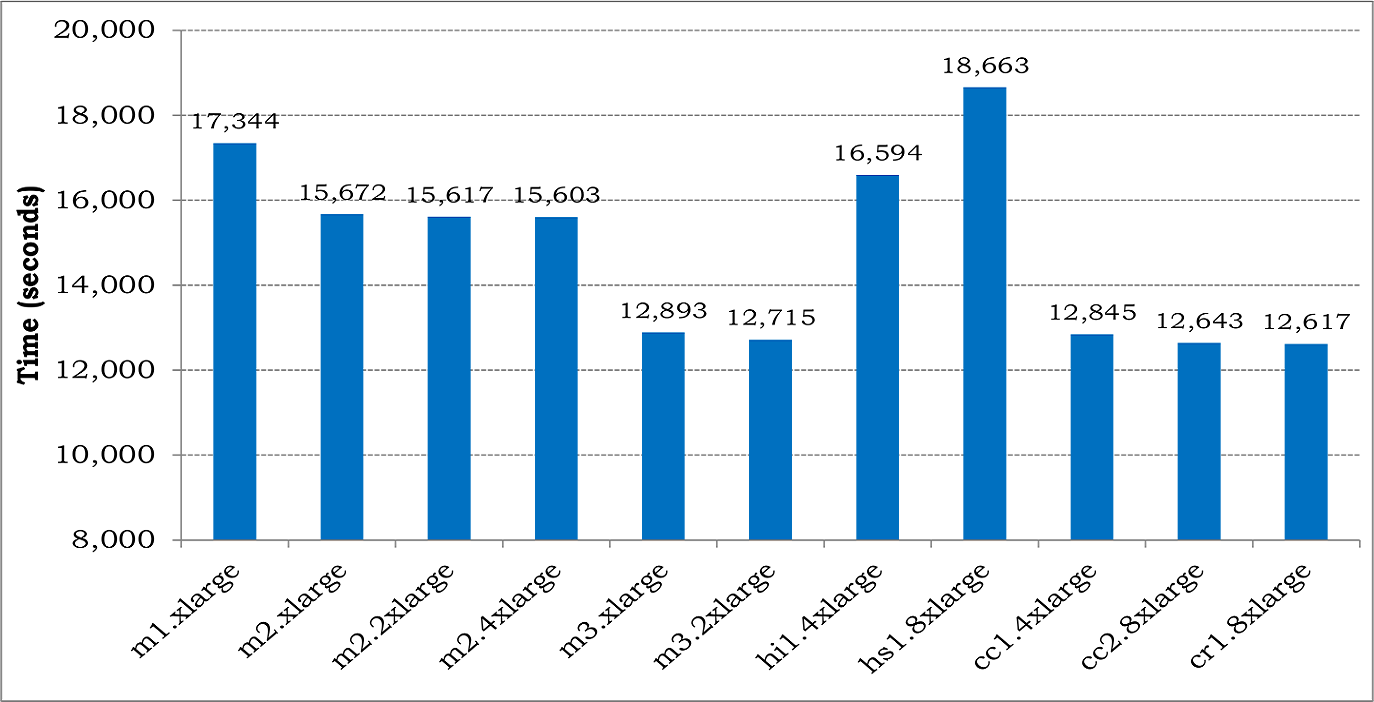} \label{figure5-1}}\hfill
\subfigure[Parallel performance on 32 threads]{\includegraphics[width=0.44\textwidth]{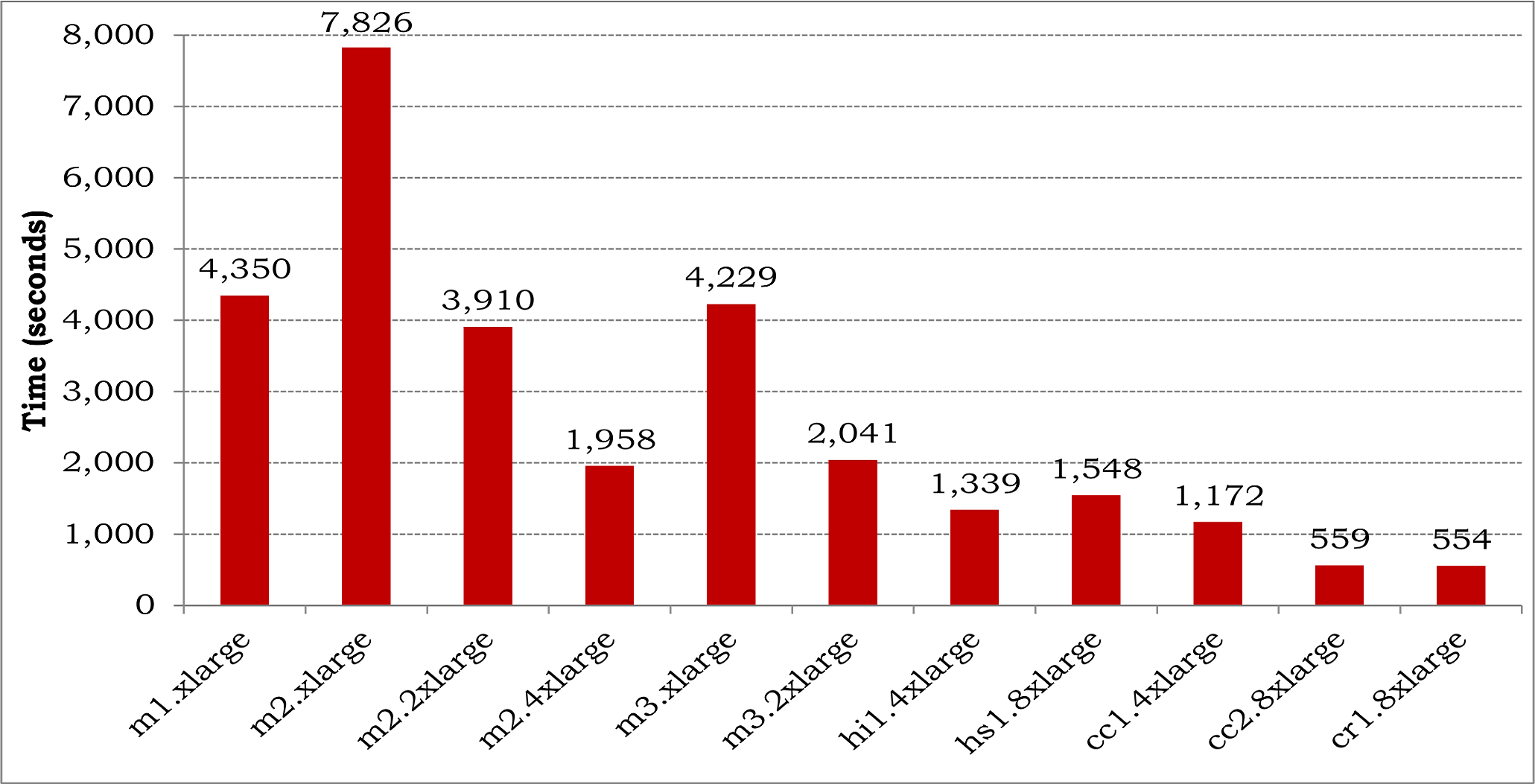} \label{figure5-2}}
\caption{Empirical study of the molecular dynamics case study}
\end{center}
\vspace{-1\baselineskip}
\end{figure}

Figure \ref{figure5-1} and Figure \ref{figure5-2} shows the time taken by the molecular dynamics simulation sequentially (on one thread) and in parallel (on 32 threads). Sequential performance varies from a minimum of nearly 12,500 seconds on the \texttt{m3}, \texttt{cc2} and \texttt{cr1} instances to a maximum of over 18,600 seconds on the \texttt{hs1} instance; there is a 50\% difference in the baseline performance. Parallel performance is significantly different; 550 seconds on \texttt{cr1} and \texttt{cc2} versus 7825 seconds on \texttt{m2.xlarge} yielding a 14x speed up.

\subsection{Comparing the Cloud Rankings}
For the first case study there is a correlation of nearly 93\% and over 60\% for sequential and parallel performance respectively between the rankings produced by the benchmarking methodology and the empirical analysis (90\% and over 71\% in the second case study). 

The key observation from the validation study is that the benchmarking methodology can consistently point to the top performing instances without the need of actually running the workload. The user can provide a set of weights based on the requirements of an application as input to the methodology to obtain a ranking of the VMs that can maximise the performance of the application.

\section{Discussion and Conclusions}
\label{conclusions}
The landscape of cloud benchmarking considers the evaluation of the resource and the service \cite{cloudbenchmark-9a}. Resource benchmarking can help in selecting VMs that can provide maximum performance when an application needs to be deployed. Service benchmarking is important to understand the reliability and variability of the service offered on the cloud \cite{cloudbenchmark-10}. We assume reasonable service is obtained on the public cloud and hence service benchmarking is not considered in this paper. 

Many of the cloud benchmarking techniques rely on data from the provider \cite{cloudbenchmark-90}. In addition they are limited in two ways. Firstly, the requirements of applications that need to be deployed on the cloud are seldom taken into account \cite{cloudbenchmark-10}. In this paper, the methodology considers user assigned weights that describe the requirements of an application along with data obtained from benchmarking VMs.

Secondly, benchmarking techniques do not focus on incorporating methods to validate the benchmarks \cite{cloudbenchmark-9a, cloudbenchmark-3}. This is necessary to guarantee that the benchmarks obtained are acceptable. Empirical analysis can be easily used for validating benchmarks but is not employed on the cloud \cite{cloudbenchmark-19, cloudbenchmark-20a}. In this paper, a validation technique empirically verifies the benchmarks using case study applications.

The research in this paper was motivated towards addressing the question: how can applications be deployed on the cloud to achieve maximum performance? We hypothesized that by taking into account the requirements of an application, along with cloud benchmarking data, VMs can be ranked in order of performance. A benchmarking methodology was developed, which takes the requirements of an application in the form of a set of user provided weights to produce a performance-based ranking of the VMs. The methodology maps the weights onto groups of attributes that describe VMs; with the view that the top ranked VMs will provide maximum performance. 

To validate our hypothesis, we performed an empirical analysis using two case study applications employed in real world that are representative of workloads which can benefit from the cloud. The key result is given a set of weights the methodology can point to the best performing VMs without having to execute the workload. In the future, we aim to incorporate a cost model that can determine the value-for-money VM for a given workload.

\end{document}